\documentclass[a4paper,aps,pre,twocolumn,groupedaddress,showkeys,showpacs,floats,floatats,floatfix]{revtex4-1}
\usepackage[utf8]{inputenc}
\usepackage[english]{babel}
\usepackage{graphicx}
\usepackage{dcolumn} 
\usepackage{bm}
\usepackage{natbib}
\usepackage{latexsym}
\usepackage{mathrsfs}
\usepackage{amssymb}
\usepackage{amsmath}
\usepackage{mathtools}
\usepackage{amscd}
\usepackage{color}
\usepackage{pifont}
\usepackage{pstricks,pst-node,pst-text,pst-3d}
\usepackage{verbatim}
\usepackage{ulem}
\usepackage[T1]{fontenc}
\usepackage{hyperref}
\hypersetup{
  colorlinks   = true, 
  urlcolor     = magenta, 
  linkcolor    = red, 
  citecolor    = blue 
}
\newrgbcolor{Red}{1.0 0.0 1.0}
\begin{document}
\title{Proliferation of stability in phase and parameter spaces of
  nonlinear systems} 
\author{Cesar Manchein$^{1}$}
\email{cesar.manchein@udesc.br}
\author{Rafael M. da Silva$^{2}$}
\email{rmarques@fisica.ufpr.br}
\author{Marcus W. Beims$^{2}$}
\email{mbeims@fisica.ufpr.br}
\affiliation{$^1$Departamento de F\'\i sica, Universidade do Estado de 
  Santa Catarina, 89219-710 Joinville, SC, Brazil} 
\affiliation{$^2$Departamento de F\'\i sica, Universidade Federal do
  Paran\'a, 81531-980 Curitiba, PR, Brazil}
\date{\today}
%
\begin{abstract}
  In this work we show how the composition of maps allows us to
  multiply, enlarge and move stable domains in phase and parameter
  spaces of discrete nonlinear systems. Using H\'enon maps with distinct
  parameters we generate many identical copies of isoperiodic stable
  structures (ISSs) in the parameter space and attractors in phase
  space. The equivalence of the identical ISSs is checked by the largest
  Lyapunov exponent analysis and the multiplied basins of attraction
  become riddled. Our proliferation procedure should be applicable to
  any two-dimensional nonlinear system. 
\end{abstract}
%

%
\maketitle

{\bf
In high-dimensional dynamical systems the possibility of controlling the 
dynamics through parametric changes is of great interest. Adding a
time dependent parameter it is possible to control the dynamics of the
paradigmatic H\'enon map generating multiply Isoperiodic Stable
Structures (ISSs) on the parameter space and consequently increasing
the number of attractors on the phase space. Numerical simulations and
analytical results explain the origin of new stable domains due to
saddle-node bifurcations for specific parametric combinations. The
distance between the multiplied ISSs can be controlled by the
intensity of the time dependent parameter and general rules for the
occurrence of proliferation are treated in details. We believe that
the present study represents a significantly new insight on the use of
alternating forces to control the dynamics of complex nonlinear
systems modeled by two-dimensional maps and can be extended to various
applications ranging from physics, biology to engineering. 
}

\section{Introduction}
\label{introd}
A huge number of physical systems in nature present regular or chaotic
dynamics depending on parameters and initial conditions. We mention
granular dynamics, coupled networks, brain dynamics, market crisis, chemical 
systems, laser physics, transport, extreme events, weather forecast, among 
many others. In nonlinear dynamics it is essential to know the correct
parameter combination which leads to (or avoids) a specific dynamics. One 
crucial step for this was the discovery of structures in the parameter 
space of {dynamical systems.} Such structures, called Isoperiodic
Stable Structures (ISSs), are Lyapunov stable islands in the parameter
space, and are supposed to be  generic in dynamical systems. For 
parameters chosen inside the ISSs the corresponding dynamics is stable and 
regular. ISSs were found in many systems, and we would like to mention some 
of them. In theoretical \cite{gallasCSF13} and experimental \cite{stoop10}
electronic circuits, continuous systems \cite{kapral82,broer98,bonatto05,
kurths06,bonattoR07,rene11,gon13}, maps \cite{kapral82,markus89,mira91,jasonPRL93,
denis11,diego11,rene16} lasers {models} \cite{gallas-diode10}, cancer models
\cite{gallas-gallas-gallas}, classical \cite{alan11-1,alan11-2,alan13-1} 
and quantum ratchet systems \cite{carlo12,alan15,carlo16}. For the
description of nature {processes} it is essential to discover generic
properties for parameter combinations in nonlinear dynamical systems
which can be applied  to {\it any} realistic situation, independent
of the specific physical  system.

In this work we investigate the not trivial dynamics of the 
composition of two-dimensional discrete maps. We use the paradigmatic 
H\'enon map (HM), whose relevant dynamics should be visible in any 
two-dimensional dissipative map. {It is  shown} that composing
HMs with distinct parameters, following a specific protocol, it is possible 
to  generate multiple ISSs which can be split in the parameter space. 
Multioverlapping identical copies of the ISSs start to separate from each 
other with increasing intensity of the perturbative parameter $F$, enlarging 
the available stable domain in phase and parameter spaces. 
The generated overlapping ISSs are {\it enlarged} ISSs, found to be the 
factorized composition of identical copies of the original ISSs. The 
proposed method is generic and can be applied to ordinary problems 
involving nonlinear behaviors. Results for du-, tri-, sextu- and 
decuplications are described for the composition of H\'enon maps with 
distinct parameters. 
Indications for the possible duplication of structures in parameter 
space was given for the composition of two quadratic coupled maps in the 
context of chaos suppression \cite{kurths96}. The replication of a 
shrimp-like ISSs was observed in a continuous oscillator \cite{rene11}, 
but its origin remained unknown. This work extends previous 
results for one-dimensional systems \cite{rafael17-1} to the non-trivial 
two-dimensional case.

The paper is presented as follows. In Sec.~\ref{HM} {we summarize the main 
properties observed in the one-dimensional case and in Sec.~\ref{2D}} the 
proliferation of {shrimp-like} ISSs in the parameter space of the two-dimensional
H\'enon map is presented. {Section \ref{attractor} shows that multiple 
attractors are created in phase space with the corresponding riddled 
basin of attraction.} {In Sec.~\ref{gen} we generalize our procedure
showing the duplication of other more complicated ISSs and Sec.~\ref{ana} 
shows analytical results for the duplication} of period $2$ (shortly 
written per-$2$) {stability boundaries in parameter space.
This corresponds to the duplication of ISSs.}
Section \ref{conclusions} summarizes our results. 

\section{Rules from The one-dimensional case}
\label{HM}
Recently it was shown \cite{rafael17-1} that by controlling the
dynamics of composed one-dimensional quadratic maps (QMs), multiple
independent attractors and independent shifted bifurcation diagrams
{can be generated. The appearance of extra stable motion 
together with the prohibition of period doubling bifurcations (PDBs) 
is the mechanism which leads to shifted bifurcations diagrams}. An 
analogous mechanism was revealed many years ago \cite{eno03} in the 
context of taming chaos in continuous systems under weak harmonic 
perturbation. 
{The above mentioned mechanism can be briefly explained in 
a simple example. Consider the} modified Quadratic Map (MQM) 
$x_{n+1} = a-\,x_n^2 + F\,(-1)^n$, with $n=0,1,2,\ldots,N$, and parameters 
($a,F$), {where $F$ is the intensity of the 
external force with alternating signal $+F,-F,+F,-F,\ldots$. Note that this 
is a composition of two QMs with alternating ($k=2$ periodic) parameters.
Besides the period $k$ of the external force, we have also the period $p$
of the variable $x_n$. For $F=0$ the above map suffers a PDB from period 
$p=1\to 2$ at $a_{1\to2}=0.75$ and a PDB from period $p=2\to 4$ 
at $a_{2\to4}=1.25$. It is clear that for $F\ne0$ no orbit with period 
$p=1$ exists anymore and the PDB at $a_{1\to2}$ becomes forbidden. 
In fact, it was shown \cite{rafael17-1} that this PDB is transformed in a 
saddle-node bifurcation and {two 
orbits of period $p=2$} exist with distinct stabilities. Consequently, for 
increasing values of $a$ these pair of per-2 orbits suffer PDBs at distinct 
values of $a_{2\to4}$, and subsequently along the whole PDB sequence 
$a_{4\to8}, a_{8\to16},\ldots$, leading to two independent shifted 
bifurcation diagrams.}

{Other composition of QMs can be used and in general} 
it was shown {for one-dimensional systems} that the 
$k$-composition of 
QMs with distinct parameters induces a dynamics which 
follows the rules:
{\bf (a)} generates $k$-attractors and $k$-independent shifted
  bifurcation diagrams when $\omega \in \mathbb{Z}$, where $\omega=p/k$. 
  In this case $p_{\mbox{\tiny F}}=p$, where $p_{\mbox{\tiny F}}$ is the 
  orbital period for $F\ne0$.
{\bf (b)} when $\omega \notin \mathbb{Z}$, the orbits of period $p$ 
 become $p_{\mbox{\tiny F}}$-periodic, where $p_{\mbox{\tiny F}}=k\,p$.
These apparently simple rules generate complex behaviors. For example,
suppose a PDB sequence $p\to 2p\to 4p\to\ldots, l\,p$. For $F\ne 0$, 
{\it all} PDB with $\omega=l\,p/k<1$, become forbidden by  the 
composition of the $k$ QMs. This prohibition is responsible for the 
generation of new periodic orbits via a saddle-node bifurcation and 
$k$-shifted bifurcation diagrams are created.

\section{Two-dimensional case}
\label{2D}

{While our paper \cite{rafael17-1} explains the basic mechanism for 
the appearance of multiple bifurcation diagrams in a family 
of one dimensional quadratic maps, the present work analyses the effects 
of such mechanism in two dimensional systems with two parameters. 
Specially we are interested in the behaviour of the ISSs, whose 
relevance in the description of dynamical systems was explained 
in Sec.~\ref{introd}.}
In general, above rules should be extended to two-dimensional systems 
with two parameters. A complex behavior is expected for the
ISSs as a function of the perturbation with many 
interesting {new} features, as will be discussed next.
The two-dimensional modified H\'enon map (MHM) is given by
\begin{align}
\begin{split}
\label{map} 
x_{n+1} &=\hspace{0.05cm} a-\,x_n^2 + b\,y_n + {g(F,n)}, \\
y_{n+1} &=\hspace{0.05cm} x_n,
\end{split}
\end{align}
with states ($x_n,y_n$) calculated at discrete times $n=0,1,2,\ldots,N$, 
parameters ($a,b,F$) {and the function $g(F,n)$ with period $k$
which will define the protocol of proliferation. For $b=0$, map 
(\ref{map})} reduces to the MQM from \cite{rafael17-1}. The 
MHM corresponds to the HM with a time dependent parameter 
$a_j^{\prime}=[a+{g(F,n)}]$. 
\begin{figure}[!b]
  \centering
  \includegraphics*[width=0.95\columnwidth]{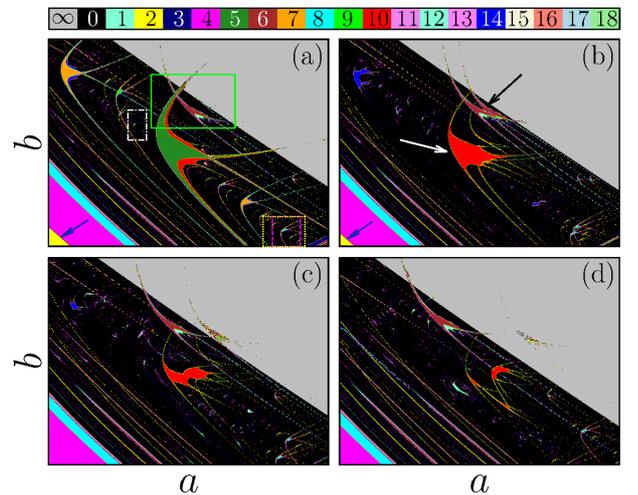}
  \caption{(Color online) Periods (see color bar) plotted in the parameter 
  space ($a,b$) inside the interval $(a_{\mbox{\tiny min}},
  a_{\mbox{\tiny max}})=(1.2,1.8)$ and $(b_{\mbox{\tiny min}},
  b_{\mbox{\tiny max}})=(0, 0.35)$, with a grid of  $10^3\times 10^3$, ICs 
  $x_0=0.01$ and $y_0=0.05$ and {$N=10^7$} iterations, for the composition 
  of $k=2$ MHMs using (a) $F=0$, (b) 
  $F=7\times 10^{-3}$, (c) $F=2\times 10^{-2}$ and (d) $F=4\times 10^{-2}$. }
  \label{henon2}
\end{figure}

\subsection{Duplication of Shrimp-like ISSs ($k=2$)}
\label{du2D}

{For the duplication we use $g(F,n) = F\,(-1)^n$ so that
the protocol is $F,-F,F,-F,\ldots$ and function $g(F,n)$ is
$k=2$ periodic.} Figure \ref{henon2} shows the period of trajectories 
in the parameter space ($a,b$). Each color represents a given period 
(see color bar). In Fig.~\ref{henon2}(a) the case $F=0$ is displayed 
and almost any ISS has shrimp-like form with distinct periods. Inside 
each ISS a sequence of PDBs $p\to 2p\to 4p\to\ldots, l\,p$ occurs. 
For a detailed description of the periods, properties of the shrimp-like 
ISSs shown in Fig.~\ref{henon2}(a), we refer the readers to the work 
\cite{jasonPRL93}. If parameters are chosen inside one ISS, the 
H\'enon map will generate a stable orbit with period corresponding to the 
color. For example, the largest ISS in the center of Fig.~\ref{henon2}(a) 
has a per-$5$, while the smaller ISS just above this per-$5$ ISS, has a 
per-$6$ (see green box). When parameters are chosen outside the ISSs, a 
chaotic motion occurs and is represented in Fig.~\ref{henon2} by the black 
color. The grey color represents the case when the trajectory diverges and 
no bounded motion is expected.

Results for $F=7\times 10^{-3}$ are shown in Fig.~\ref{henon2}(b), which 
displays the same parameter interval from Fig.~\ref{henon2}(a). For 
simplicity we have chosen to count the periods including all 
states. In other words, we do not display the periods of the 
composed map. First observation  in Fig.~\ref{henon2}(b) is that while 
ISSs with even periods keep 
their period, for all ISSs with odd periods the period is duplicated, as
expected by rule {\bf (b)}. For example, the large shrimp-like ISS 
mentioned above with $p=5$ ($\omega=5/2$), has now $p_{\mbox{\tiny F}}=10$ 
(see white arrow). There are obviously no per-$1$ orbits anymore and orbits 
with odd periods-$p$ are prohibited since $\omega=p/k=p/2$ is 
rational.  The second observation is that all ISSs with even periods $p$ 
start to {\it duplicate}, since $\omega=p/2$ is an integer and satisfies 
rule {\bf (a)}.  This is better observed by the ISSs with periods
$p=p_{\mbox{\tiny F}}=6$ ($\omega=6/2$), which separate from each other. 
See the black arrow indicating both ISSs. From the resolution of 
Fig.~\ref{henon2}(b), some of the duplications from other ISSs cannot 
be seen. The separation between the duplicated ISSs in the parameter 
space increases with $F$, as can be checked in 
Fig.~\ref{henon2}(c) for $F=2\times 10^{-2}$ and in Fig.~\ref{henon2}(d) 
for $F=4\times 10^{-2}$. An interesting aspect is that the per-$6$ ISS 
from the right moves further to the right as $F$ increases, until it reaches 
the grey region where it starts to disappear. In addition, the above mentioned
large shrimp-like ISS with $p_{\mbox{\tiny F}}=10$ (white arrow) is 
transformed into three (not a triplication in this case, see explanation 
bellow) interconnected shrimps observed in Fig.~\ref{henon2}(d). Such 
interconnected shrimps were observed to be relevant in a tunnel diode and a 
fiber-ring laser \cite{francke13} {and, in this context, endorses the 
importance to control the intermediate dynamics to create and enlarge the 
ISSs.}
\begin{figure}[!t]
  \centering
  \includegraphics*[width=0.9\columnwidth]{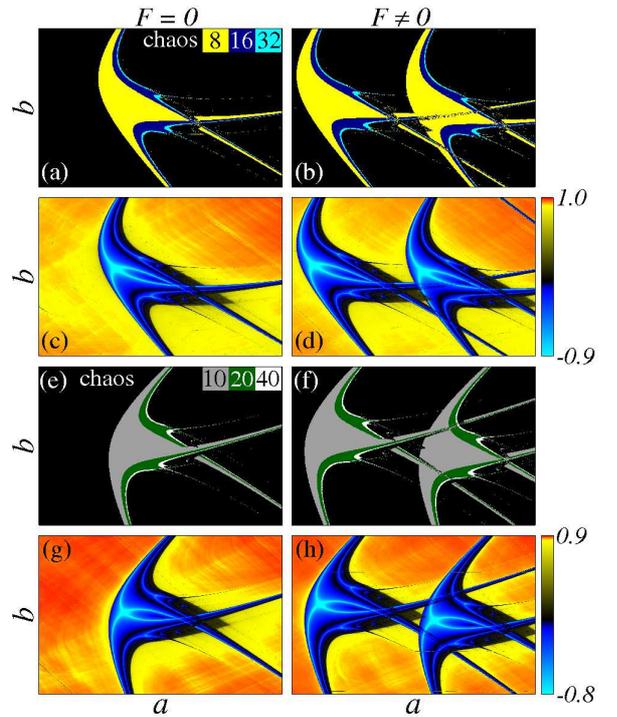}
  \caption{(Color online) Parameter space $(a,b)$ with a grid of 
  $10^3\times 10^3$. The duplication of a per-8 ISS is presented in 
  (a)-(d) in the interval $(a_{\mbox{\tiny min}}, a_{\mbox{\tiny max}})
  = (1.719,1.736)$, $(b_{\mbox{\tiny min}}, b_{\mbox{\tiny max}}) = 
  (0.110,0.118)$. Periods were count after a transient of $5 \times 10^{6}$ 
  iterations and are shown for (a) $F=0$ and (b) $F=2\times 10^{-3}$. 
  The largest LE was computed using a trajectory of $5\times 10^{6}$ 
  iterations and is displayed in (c) and (d) for same values of $F$. 
  For a per-10 ISS in the interval $(a_{\mbox{\tiny min}}, 
  a_{\mbox{\tiny max}}) = (1.6726,1.6794)$, $(b_{\mbox{\tiny min}}, 
  b_{\mbox{\tiny max}}) = (0.1417,0.1441)$ the duplication is presented in (e)
  $F=0$ and (f) $F=8\times10^{-4}$ using periods and in (g) and (h) using 
  the largest LE.}
  \label{henon-lyap}
\end{figure}
%
\subsubsection{Magnification of ISSs with integer $\omega=p/k$}
To understand better the duplication of ISSs, Fig.~\ref{henon-lyap} presents
details of the effect of increasing values of $F$ in two distinct 
shrimp-like ISSs with PDBs $p=p_{\mbox{\tiny F}}=8\to 16\to32\ldots$
($\omega=4\to 8\to 16\ldots$) in Figs.~\ref{henon-lyap}(a) and (b) and 
$p=p_{\mbox{\tiny F}}=10\to20\to40\ldots$ ($\omega=5\to10\to20\ldots$)
in Figs.~\ref{henon-lyap}(e) and (f). For $F\ne0$ in both cases we 
{obtain $k=2$ identical ISSs which are copies of the original 
one's and $k=2$ attractors in phase space (see Sec.~\ref{attractor})}. 
To turn the statement of identical copies more convincing, we plot the 
largest Lyapunov exponent (LE) for each case [see 
Figs.~\ref{henon-lyap}(c),(d),(g),(h)].
Grey, yellow to red for increasing positive LE, and blue to cyan for 
increasing negative LE. It nicely shows that the internal structures of 
the ISSs, which contain information about the local stability, is 
unaltered by the duplication. Thus, identical copies refer to the
shape of the ISSs and the corresponding stability for parameters chosen 
inside the ISSs.

\subsubsection{Magnification of ISSs with rational $\omega=p/k$}
Figure \ref{henon-p7} presents the magnification for a shrimp-like ISS 
with PDB $p=7\to14\to26\to\ldots$ ($\omega=7/2\to7\to14\dots$). Here 
the effect of increasing values of $F$ on the ISS is more  complicated 
since the lowest period of the PDB sequence follows rule {\bf (b)}, while 
all subsequent periods follow rule {\bf (a)}. Since $\omega=7/2$ for 
$p=7$, we obtain $p_{\mbox{\tiny F}}=2\times 7=14$, confirming rule {\bf 
(b)}. In this case only one attractor with $p_{\mbox{\tiny F}}=14$ is 
found and no duplication {of the ISS with this period} occurs.
This breaks the ISS apart as observed in Figs.~\ref{henon-p7}(b)-(d). 
While one main shrimp-like ISSs with period $p_{\mbox{\tiny F}}=14$ 
(and PDBs sequence $28\to 56\ldots$) remains, two smaller 
{non-overlapping} ISSs with period $p=p_{\mbox{\tiny F}}=14$ 
move apart (see white arrows). However, here 
the internal structure of the ISS is changed qualitatively when 
compared to the $F=0$ case, as can be checked in the largest LE 
analysis in Figs.~\ref{henon-p7}(e)-(h). 
\begin{widetext}
$\quad$
\begin{figure}[!t]
  \centering
  \includegraphics*[width=0.9\columnwidth]{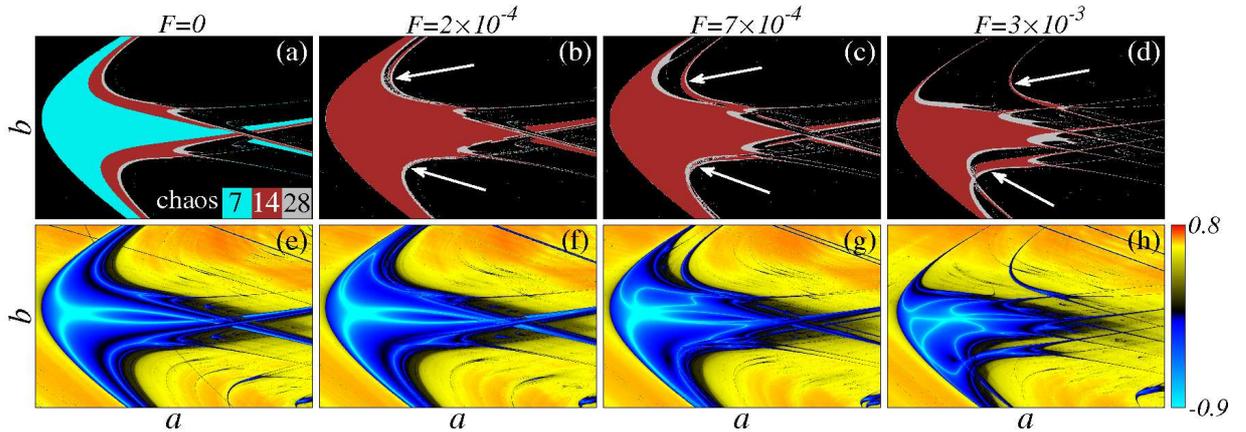}
  \caption{(Color online) Parameter space $(a,b)$ with a grid of $10^3 
  \times 10^3$ showing the interval $(a_{\mbox{\tiny min}}, a_{\mbox{\tiny max}})
  = (1.225,1.295)$, $(b_{\mbox{\tiny min}}, b_{\mbox{\tiny max}}) = 
 (0.26,0.33)$ displaying  in (a)-(d) the periods determined after a 
transient time of $5 \times 10^6$  in  (e)-(h) the LE computed using a 
 trajectory of $5\times 10^6$ iterations.}
  \label{henon-p7}
\end{figure}
\end{widetext}

\subsection{Triplication ($k=3$), quadruplication ($k=4$) and more...}
\begin{figure}[!h]
  \centering
  \includegraphics*[width=0.9\columnwidth]{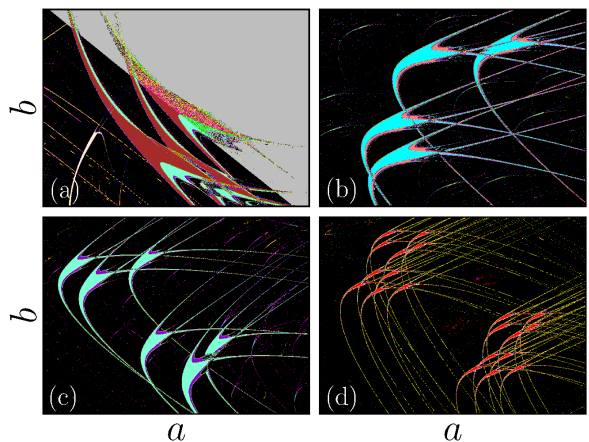}
  \caption{(Color online) Periods (see color bar in Fig.~\ref{henon2}) 
  plotted in the parameter space ($a,b$) with a grid of $10^{3}\times 10^3$  
  and $N=2\times 10^6$ iterations for (a) triplication ($k=3$) of the per-$6$ 
  shrimp [see green box from Fig.~\ref{henon2}(a)] inside the interval 
  $(a_{\mbox{\tiny min}},a_{\mbox{\tiny max}})=(1.42,1.60)$ and 
  $(b_{\mbox{\tiny min}}, b_{\mbox{\tiny max}})=(0.20,0.29)$ using $F=1.2 \times
  10^{-3}$, (b) quadruplication ($k=4$) of the per-$8$ shrimp [see magenta 
  box from 
  Fig.~\ref{henon2}(a)] for $F=2.0\times 10^{-3}$ and $(a_{\mbox{\tiny min}},
  a_{\mbox{\tiny max}})={(1.68,1.74)}$, $(b_{\mbox{\tiny min}}, b_{\mbox{\tiny max}})=
  {(0,0.05)}$, (c) sextuplication ($k=6$) of the per-$12$ shrimp 
  [see white box from Fig.~\ref{henon2}(a)] for $F=1.2\times 10^{-3}$, 
  $(a_{\mbox{\tiny min}},  a_{\mbox{\tiny max}})={(1.37,1.41)}$, $(b_{\mbox{\tiny min}}, 
  b_{\mbox{\tiny max}})={(0.18,0.23)}$ and (d) examples of the decuplications 
   ($k=10$) of per-$10$ of two shrimp-like ISSs in the interval 
  $(a_{\mbox{\tiny min}},a_{\mbox{\tiny max}})=(1.66,1.75)$,
  $(b_{\mbox{\tiny min}},b_{\mbox{\tiny max}})=(0,0.05)$ {[see yellow box 
  from Fig.~\ref{henon2}(a)]}.}
  \label{henon3}
\end{figure}

Next it is shown that the above behavior can be extended to multiply 
ISSs in the parameter space. For the triplicated case ($k=3$) the external 
force must have per-$3$, as can be obtained by using the protocol
$-F,0,F,-F,0,F,\ldots$ perturbing the HM. In this case, 
the ISSs with periods multiple of $3$ are triplicated. In this section we 
focus on integer values of $\omega$. Results are shown in Fig.~\ref{henon3}(a), 
which displays a magnification of the parameter space from Fig.~\ref{henon2}(a) 
and for $F=1.2\times 10^{-3}$. This is a triplication of the per-$6$ shrimp 
($\omega=6/3$). In fact, it creates $k=3$ per-$6$ stable periodic orbits 
which separate more and more for increasing values of $F$. In Fig.~\ref{henon3}(b) 
it is shown the case of the quadruplication of the per-$8$ shrimp [see
magenta box from Fig.~\ref{henon2}(a)] using $+F,-F/2,F/2,-F,\ldots,+F,-F/2$ 
with $F=2.0\times 10^{-3}$. Fig.\ref{henon3}(c) displays the sextuplication of the 
per-$12$ shrimp using $+F,-F/2,F/4,-F/4,F/2,-F,+F,-F/2,\ldots$ with $F=1.2\times 10^{-3}$. 
To show that a proliferation of the ISS is possible, we present in Fig.~\ref{henon3}(d) 
the case of the decuplications of {\it two} per-10 ISSs, so that twenty ISSs are observed. 

\section{Multiplication of attractors and riddled basins}
\label{attractor}

\begin{figure}[!h]
  \centering
  \includegraphics*[width=0.96\columnwidth]{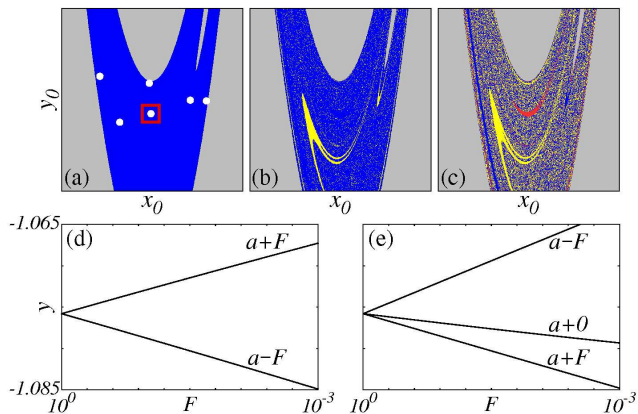}
  \caption{(Color online) The basin of attraction of the MHM plotted 
  inside  the interval $(x_{0_{\mbox{\tiny min}}}, x_{0_{\mbox{\tiny max}}}) 
  = (-3.0, 3.0)$ and $(y_{0_{\mbox{\tiny min}}}, y_{0_{\mbox{\tiny max}}}) = 
  (-7.0, 7.0)$ for $a=1.514$, $b=0.230$. In (a) we show the case $F=0$ and the 
  white circles are the periodic points of the per-6 attractor. In (b)
  the duplication ($k=2$) for  $F=1\times 10^{-3}$ and (c) the triplication
  ($k=3$) for $F=1\times 10^{-3}$. The gray color represents the ICs that lead 
  to divergence and the regions with blue, yellow and red colors refer to 
  different attractors due to $F\neq 0$. In (d) and (e) we show, respectively, 
  the duplication and triplication of the orbital point from the per-6 attractor 
  inside the red box in (a).}
 \label{phase}
\end{figure}

It remains to show that the multiplication of ISSs in the parameter 
space is a consequence of the multiplication of attractors in phase space. 
To exemplify this we show the duplication and triplication of 
shrimp-like ISSs from Fig.~\ref{henon2}. Figure \ref{phase}(a) shows 
the basin of attraction inside the per-6 ISS from Fig.~\ref{henon2}(a) 
for $F=0$. Figure \ref{phase}(b) shows the basin of attraction for the 
duplication for which the identical copies of the ISSs {still} 
overlap. We clearly 
observe that it generates another basin of attraction  related to the 
duplicated per-6 orbit. Figure \ref{phase}(c) shows the basin of 
attraction for the case of three attractors, each one with per-6. 
%

Figures \ref{phase}(d) and (e) display, respectively, the duplication and 
triplication of one per-6 orbital point [see inside red box in Fig.~\ref{phase}(a)] 
as a function of $F$. It is very interesting to observe that $k$ small parametric 
changes in the HM generates $k$ riddled basin of attractions \cite{yorke92}.


\section{Duplication of other Structures}
\label{gen}

\begin{figure}[!b]
  \centering
  \includegraphics*[width=0.9\columnwidth]{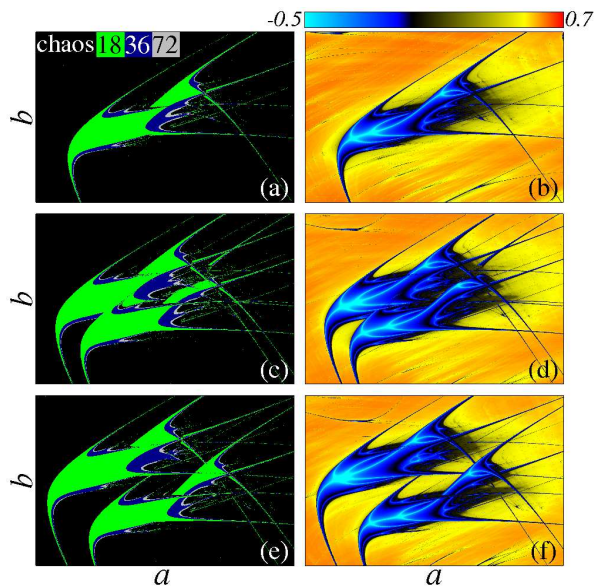}
  \caption{(Color online) Parameter space $(a,b)$ with a grid of 
  $10^3\times 10^3$. The duplication of a per-18 ISS is presented  
  in the interval $(a_{\mbox{\tiny min}}, a_{\mbox{\tiny max}})
  = (1.289,1.302)$, $(b_{\mbox{\tiny min}}, b_{\mbox{\tiny max}}) = 
  (0.259,0.268)$ displaying the periods in (a) $F=0$, (c) 
  $F=6\times 10^{-4}$  and (e) $F=1\times 10^{-3}$ and the largest LE 
  in (b), (d) and (f) for same values of $F$.}
  \label{p18}
\end{figure}

The ISSs discussed in Sec.~\ref{du2D} have the well known 
shrimp-like form \cite{jasonPRL93}. However, other ISSs exist which 
may be more complicated or not. For example, simpler ISSs than the
shrimp-like are the cuspidal and non-cuspidal, shown respectively in
Fig.~1(a) and  Fig.~1(b) from \cite{alan11-2}. More complicated and 
higher order ISSs were described in the very last paper from Lorenz 
\cite{lorenz08}.

The purpose of the present Section is to show that our multiplication 
procedure is also valid for such ISSs.
The first example is shown in Fig.~\ref{p18} for the duplication
of a per-18 ISS. Doing a visual analysis, the ISS from Fig.~\ref{p18}(a) 
for $F=0$ could be wrongly interpreted as a composition of shrimp-like ISSs 
which are overlapped. But this is not
the case, as shown by Lorenz and checked here analyzing the LE in 
Fig.~\ref{p18}(b). Compared to the shrimp-like ISSs, now we have two
superstable regimes (cyan lines) inside the ISSs. This also suggest
that the dynamics inside the ISS from Fig.~\ref{p18}(a) is different, 
regarding stability, from the dynamics inside shrimp-like ISS.

As the values of $F$ increases the duplication of the ISS is 
visible and nice complex pictures are generated. While the inner
structure of the LE inside the ISSs from Fig.~\ref{p18}(d) are still 
identical to the original one from Fig.~\ref{p18}(b) (compare cyan 
lines), this changes in Fig.~\ref{p18}(f).  We observed in general 
that the higher-order ISSs are more sensitive to $F$. 
In other words, the duplication generates identical copies of the
higher-order ISS, but they change very fast with increasing values of 
$F$.

\begin{figure}[!t]
  \centering
  \includegraphics*[width=0.9\columnwidth]{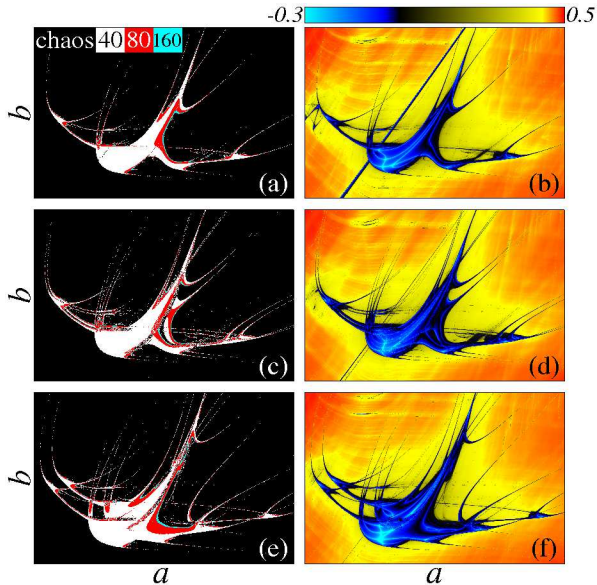}
  \caption{(Color online) Parameter space $(a,b)$ with a grid of 
  $10^3\times 10^3$. The duplication of a per-40 ISS is presented in 
  the interval $(a_{\mbox{\tiny min}}, a_{\mbox{\tiny max}})
  = (1.1703,1.1722)$, $(b_{\mbox{\tiny min}}, b_{\mbox{\tiny max}}) = 
  (0.3420,0.34235)$. Cases (a) $F=0$, (c) $F=2\times 10^{-6}$
  and (e) $F=1\times 10^{-5}$ display the periods and 
  (b), (d) and (f) the largest LE for same values of $F$.}
  \label{p40}
\end{figure}

Figure \ref{p40} shows the example of the duplication of a 
per-40 higher-order ISS. Again the duplications are visible but the
copies are only identical for very small values of $F$. Both 
examples above are related to integer values of $\omega=p/k$, namely 
$18/2=9$ and $40/2=20$ respectively. Both cases obey rule {\bf (a)}. 

\section{Analytical results for $p_{\mbox{\tiny F}}=2$}
\label{ana}

For low periods it is possible to give an analytical demonstration of the 
duplication. Years ago  the boundaries between the born of per-$1$ and the 
PDBs $1\to 2$ and $2\to 4$ were determined analytically in the parameter 
space of the H\'enon map. These boundaries are given by the relations 
\cite{beims-gallas97,gallas95}:
\begin{eqnarray}
W_{1}(a,b)&=& (4a+1-2b+b^2)=0,\cr
& & \cr
W_{1\to 2} (a,b)&= & (4a-3+6b-3b^2)^3=0,  \nonumber
\end{eqnarray}
\vspace*{-0.6cm}
\begin{align}
 W_{2\to4}& (a,b) = (4a-5+6b-5b^2)^2\times [5b^4 +4b^3+ \cr
 & (8a-2)b^2+
(16a+4)b +16a^2 +8a +5]\, =0.\nonumber
\end{align}
Compared to the original work, last equation includes a polynomial 
with complex solutions for ($a,b$). This polynomial must be taken into 
account in case $F\ne 0$. Applying the same procedure for the 
{duplication in} MHM we obtain
%
\begin{align}
\label{eq1-mhm}
&W_{p_{\mbox{\tiny F}}=2} (a,b,F) = W_{1}(a,b) W_{1\to 2} (a,b) + 256F^4-
 \cr
&\qquad[288b^4 -1152b^3 +(1536a+1728)b^2 -(3072a+\cr
&\qquad1152)b + 512a^2 +1536a +288]F^2=0,\\
%
\label{eq2-mhm}
&\cr
&W_{p_{\mbox{\tiny F}}=2\to4}(a,b,F) = W_{2\to4} (a,b)+ 256F^4 -[160b^4-\cr
&\qquad 1152b^3 +(1536a +1472)b^2 - (3072a +1152)b + \cr
&\qquad 512a^2 +1536a +160]F^2=0.\\
\nonumber
\end{align}
%
Equation (\ref{eq1-mhm}) gives the birth (saddle-node bifurcation)
of per-1 orbits for the map composed of two iterations of the MHM
($p_{\mbox{\tiny F}}=2$) and Eq.~(\ref{eq2-mhm}) indicates the PDB
from period 1 to 2 for the composed map, which for the iterations of
MHMs means $p_{\mbox{\tiny F}}=2\to 4$. {The solution of 
Eq.~(\ref{eq2-mhm}) is indicated with blue arrows in Fig.~\ref{henon2}(a)
for $F=0$ and in Fig.~\ref{henon2}(b) for $F=7\times10^{-3}$.} 
Interesting to observe that
for $F\ne 0$ the boundaries $W_1 (a,b)$ and $W_{1\to 2} (a,b)$ become 
{\it coupled} in Eq.~(\ref{eq1-mhm}). In other words, the two 
independent conditions $W_1 (a,b)=0$ for saddle-node bifurcation and 
$W_{1\to 2} (a,b)=0$ for PDB, are transformed in {\it one} saddle-node 
bifurcation condition $W_{p_{\mbox{\tiny F}}=2} (a,b,F)=0$. Therefore
the PDB $1\to2$ from $F=0$ becomes forbidden. 

To explain {this better} we show an example using $b=0.3$ and 
$F=0.01$. The solutions for the first boundaries and $F=0$ are
\begin{equation}
 W_1 (a,0.3): a=-0.1225,\quad W_{1\to 2}(a,0.3): a=0.3675,
\nonumber
\end{equation}
while the solutions for $F=0.01$ become 
\begin{align}
& W_{p_{\mbox{\tiny F}}=2}(a,0.3,0.01): a=-0.1224\quad \mbox{and}\quad 
a=0.4377.
\nonumber
\end{align}
This shows that the birth of ${p_{\mbox{\tiny F}}=2}$ is shifted to the left 
($-0.1225\to-0.1224$) and the PDB $1\to2$ at  $a=0.3675$ becomes 
forbidden (since $\omega=1/2$), transforming it into a saddle-node 
bifurcation at $a=0.4377$. Thus, for $F=0.01$ we have two saddle-node 
bifurcating points. The other boundaries are given by
\begin{eqnarray}
& & W_{2\to4}(a,0.3): a=0.9125\,\, \mbox{and}\,\,a=-0.4225\pm i\,0.4550,\cr
& &\cr
& &W_{p_{\mbox{\tiny F}}=2\to4}(a,0.3,0.01): a=0.8983\,\, \mbox{and}\,\,  
a=0.9267,\nonumber
\end{eqnarray}
which shows that the complex solution from the $F=0$ case becomes real and 
we end up wit {\it two} PDBs $2\to 4$, one in $a=0.8983$ and the other one
in $a=0.9267$. This explains the origin of the duplication of the PDBs
sequence and of the ISSs which contain them. For more simples examples
of the origin of shifted bifurcation diagrams via prohibition of PDBs
we refer the reader to the one-dimensional case \cite{rafael17-1}.

\section{Conclusions} 
\label{conclusions}

In this work we show that the parametric control in composed maps can 
be used to enlarge stable domains in phase and parameter spaces of 
{two-dimensional discrete} nonlinear dynamical systems. Since 
the stable domains in parameter space are generic, our results are expected 
to be applicable to a large number of systems. We present analytical and 
numerical results for the specific case of the composition of H\'enon maps 
with distinct parameters. Using the composition of $k$ H\'enon maps with 
distinct parameters we have observed following properties:
{\bf (1)} When the ratio $\omega=p/k$ is an integer, where $p$ is the 
period of the stable orbit, $k$-identical attractors in phase space and 
$k$-identical ISSs in parameters space are generated. The identical 
copies are split apart as a function of the parameter $F$. The 
equivalence between identical stable attractors and identical ISSs was 
checked by the largest LE analysis. Besides that, the additional 
basin of attraction regarding to the identical copies of the ISSs are riddled.
{\bf (2)} When the ratio $\omega$ is not an integer, the number of 
attractors in phase space and ISSs in parameter space remain unaltered.
The new orbital period is $p_{\mbox{\tiny F}}=k\,p$ and the multiplied
ISS is broken apart.
{\bf (3)} {The sign of the parameters from the intermediate 
dynamics must change by each iteration, otherwise no multiplication
is observed.}

The multiple composition of maps lead to the 
appearance of multiple attractors in phase space and multiple shifted ISSs 
in the parameter space. Consequently occurs a considerable enlargement of 
the stable domains in phase and parameter spaces.  This is crucial for the 
survival of the desired dynamics under noise and temperature effects, which 
usually destroy the ISSs starting from their borders \cite{alan13-1} 
{(also observed in the parameter space of the relativistic standard map 
\cite{cesar17})}. Future contributions intend to verify the enlargement of 
stable domains for practical applications submitted to thermal effects.
The multiple composition of H\'enon maps may be related to the 
general {\it Jung's decomposition} \cite{jung89}, which shows that any planar, 
invertible quadratic map can be reduced to a composition of Henon-like maps.
However, it is not the  purpose of the present work to show such relation.

\acknowledgments{R.M.S. thanks CAPES (Brazil) and C.M. and M.W.B. thank 
CNPq (Brazil) for financial support. C.M. also thanks FAPESC (Brazil)
for financial support. The  authors  also  acknowledge computational 
support from Professor Carlos M.~de Carvalho at LFTC-DFis-UFPR.}


\end{document}